\documentclass[twocolumn,pra,aps,raggedbottom, superscriptaddress]{revtex4-2}
\usepackage[table,xcdraw]{xcolor}
\usepackage{amsmath,amsfonts,amssymb,graphicx,hyperref,tikz,natbib,array,braket}
\usepackage{mathptmx,textcomp,braket,multirow}
\usepackage[T1]{fontenc}
\usepackage[utf8]{inputenc}
\usepackage{multirow}
\hypersetup{colorlinks,citecolor=blue,filecolor=black,linkcolor=blue,urlcolor=blue}
  
\definecolor{lime}{HTML}{A6CE39}
\DeclareRobustCommand{\orcidicon}
{
	\begin{tikzpicture} 
	\draw[lime, fill=lime] (0,0) circle [radius=0.15] node[white] {{\fontfamily{qag}\selectfont \tiny ID}};
	\draw[white, fill=white] (-0.0625,0.095) 	circle [radius=0.007];
	\end{tikzpicture}
	\hspace{-2.2mm}
}
\newcommand\orcidID[1]{\href{https://orcid.org/#1}{\orcidicon}}

\newcommand{\be}{\begin {equation}}
\newcommand{\ee}{\end {equation}}
\newcommand{\beqa}{\begin {eqnarray}}
\newcommand{\eeqa}{\end {eqnarray}}
\newcommand{\mb}{\mathbf}

\begin{document}
 
\title{Coherent control of orthogonal continuum states in XUV photoionization}

\author{Neha Kukreti\orcidID{0009-0004-9063-0037}} 
\email{nehakukreti1111@gmail.com}
\author{Amol R. Holkundkar\orcidID{0000-0003-3889-0910}}
\email{amol@holkundkar.in}  
\affiliation{Department of Physics, \href{https://ror.org/001p3jz28}{Birla Institute of Technology and Science Pilani}, Pilani Campus, Vidya Vihar, Pilani, Rajasthan 333031, India}
 
\date{\today}

\begin{abstract}
  
We demonstrate coherent control of orthogonal continuum electron states in the XUV photoionization of atomic hydrogen using polarization-tailored driving fields. By combining polarization mixing with carrier-envelope-phase (CEP) control, we generate a pair of orthogonal momentum-space basis states whose populations and relative phase can be independently tuned via the polarization-mixing parameter and CEP, respectively. The resulting photoelectron wave packets are prepared as coherent superpositions within this effective two-dimensional subspace, with quantum coherence confirmed by interference visibilities reaching $\sim 99$\% in the photoelectron momentum distributions (PMDs). We further demonstrate that a bichromatic driving-field configuration extends this framework to a four-dimensional continuum manifold, with basis states distinguished by both angular emission patterns and radial momentum distributions, and with independent amplitude and phase control preserved across the higher-dimensional subspace. These results establish that polarization-tailored XUV fields provide a flexible route to independent amplitude and phase control within low-dimensional subspaces of the photoelectron continuum, with the engineered dynamics directly observable in momentum-resolved spectra.

\end{abstract}

\maketitle

\section{Introduction}

Photoionization and strong-field ionization provide powerful routes for generating coherent electron wave packets in the continuum~\cite{28,505,533}. Unlike bound electronic states, continuum wave packets are delocalized over a broad range of energies, momenta, and angular-momentum channels, providing access to ultrafast electron dynamics on attosecond time scales~\cite{28,533}. The resulting photoelectron momentum distributions (PMDs) encode detailed information about both the target system and the driving electromagnetic field~\cite{42,56}, making them highly sensitive observables of light-matter interactions. PMDs also provide direct access to ultrafast electron dynamics with attosecond temporal resolution~\cite{294,308}. Complementing such momentum-resolved measurements, recent experiments in photoelectron quantum-state tomography have enabled reconstruction of the photoelectron density matrix, revealing coherence and entanglement in continuum electron states~\cite{673}. Recent advances in ultrafast laser technology, including carrier-envelope-phase (CEP) stabilization~\cite{70} and the use of polarization-tailored and waveform-engineered laser fields~\cite{1,729}, have enabled unprecedented control over continuum electron dynamics, opening new opportunities for tailoring photoelectron emission in momentum space.

  More broadly, the ability to generate and coherently manipulate orthogonal continuum states may be of interest for emerging quantum technologies. Quantum information processing relies on the preparation and control of well-defined two-level systems, and a wide range of physical platforms, including trapped ions~\cite{435,364}, superconducting circuits~\cite{378,449}, neutral atoms\cite{392,463}, semiconductor spins~\cite{407,477}, and photonic systems~\cite{421,491}, have been developed for this purpose. Despite remarkable progress, different architectures offer distinct trade-offs between coherence, controllability, scalability, and 	operational speed, motivating continued exploration of alternative quantum-state platforms.  The rapid development of attosecond and polarization-controlled XUV sources has substantially expanded the degree of coherent control achievable in photoionization dynamics, suggesting that continuum-state quantum engineering may become experimentally accessible in the future. Although the present work is not directed toward quantum-information applications, it establishes key ingredients required for such possibilities by demonstrating coherent amplitude and phase control within effective continuum-state manifolds.

Considerable effort has therefore been devoted to manipulating PMDs through waveform control~\cite{687,701}, polarization engineering~\cite{715}, and quantum interference effects, including temporal wavepacket interference~\cite{42}, photoelectron holography~\cite{56}, and competing ionization pathways~\cite{168,140}. These studies have demonstrated control over emission asymmetries~\cite{168}, photoelectron angular distributions through quantum interference and channel competition~\cite{154,336}, and sub-cycle ionization dynamics~\cite{126}. Despite this progress, most investigations have focused on steering observable momentum patterns rather than establishing controllable continuum-state subspaces with well-defined quantum-state structure. A central objective is therefore to identify orthogonal continuum states that can be coherently manipulated through experimentally accessible control parameters.

Previous studies have shown that photoelectron momentum distributions encode coherent interference between electron wave packets and ionization pathways, demonstrating that continuum electrons can retain coherence and support controllable quantum interference~\cite{280,322}. These observations suggest that suitably selected continuum emission channels are effective basis states within the photoelectron continuum. If such states can be prepared, distinguished, and manipulated coherently, they provide a natural framework for describing continuum dynamics in terms of quantum superposition. In this picture, distinct momentum-space channels play a role analogous to the basis states encountered in spin, polarization, or interferometric-path systems, thereby enabling an effective low-dimensional state-space description of continuum electron dynamics.

A key challenge is therefore to determine whether continuum electron wave packets can be manipulated as effective low-dimensional quantum states. To realize such a framework, the basis states must be distinguishable and approximately orthogonal, coherent population transfer between them must be achievable through experimentally accessible control parameters, and their relative phases must be independently controllable to enable the preparation of arbitrary superposition states. In addition, the resulting dynamics must be directly observable through measurable quantities such as PMDs. Achieving these requirements remains challenging because continuum wave packets occupy a continuous spectrum of momentum states, making the identification and coherent manipulation of a finite set of well-separated orthogonal basis states considerably more difficult than in discrete-level systems. Viewed from this perspective, the realization of controllable continuum-state manifolds raises the broader question of whether continuum electron wave packets can serve as carriers of quantum information. 
 
Polarization-tailored extreme-ultraviolet (XUV) fields might provide a particularly attractive route toward addressing these challenges. Advances in the generation and control of polarized attosecond and XUV radiation~\cite{196,575,589} have created new opportunities for tailoring photoionization dynamics through the polarization and temporal structure of the driving field~\cite{112,547,561}. By combining polarization mixing with CEP control, the field trajectory can be continuously tailored, enabling selective access to distinct ionization and recollision pathways~\cite{603,617} and momentum-space emission channels~\cite{631}. Since photoelectron momentum distributions are highly sensitive to the underlying field geometry~\cite{336,168,645}, polarization shaping offers a natural mechanism for controlling both the population of continuum channels~\cite{659} and the phase structure of continuum wave packets~\cite{659,533}. These capabilities make polarization-tailored XUV fields a promising platform for constructing and steering effective continuum-state manifolds.
  
In this work, we demonstrate the generation and coherent control of orthogonal continuum states in photoelectron momentum space using polarization-tailored XUV fields. By combining polarization mixing with CEP control, we achieve independent manipulation of state amplitudes and relative phases within effective two- and four-state continuum manifolds, establishing a framework for engineering low-dimensional coherent state spaces embedded within the photoelectron continuum. The essential ingredients of coherent quantum-state engineering, preparation, manipulation, and readout of arbitrary superpositions of orthogonal basis states are each explicitly demonstrated and remain directly observable through the photoelectron momentum distributions.

The remainder of the paper is organized as follows. Section ~\ref{sec:theory} describes the theoretical and numerical methods, and Sec.~\ref{sec:results} presents the results and discussion, beginning with the two-state polarization-mixing case before extending to the four-dimensional bichromatic manifold. Moreover, Sec.~\ref{sec:conclusion} concludes the paper. Two appendices are included to rigorously establish the quantum coherence of the basis states and to characterize the independent action of the control parameters. 
 
\section{Theoretical and Numerical Considerations} \label{sec:theory}

The interaction of the XUV fields with atomic hydrogen is simulated by solving the full-dimensional time-dependent Schrödinger equation (TDSE) using the time-dependent generalized pseudospectral method in length-gauge. Although the atomic potential is described by the single-active-electron (SAE) model potential of Ref.~\cite{519} for noble-gas atoms, it reduces exactly to the pure Coulomb form \(V(r) = -1/r\) in the present case of hydrogen. Details of the numerical implementation, convergence analysis, and the procedure used to extract photoelectron momentum distributions have been reported in Refs.~\cite{1,15} and are not repeated here. The atomic units are used throughout the manuscript. 

\begin{figure}[b]
	\includegraphics[width=\linewidth]{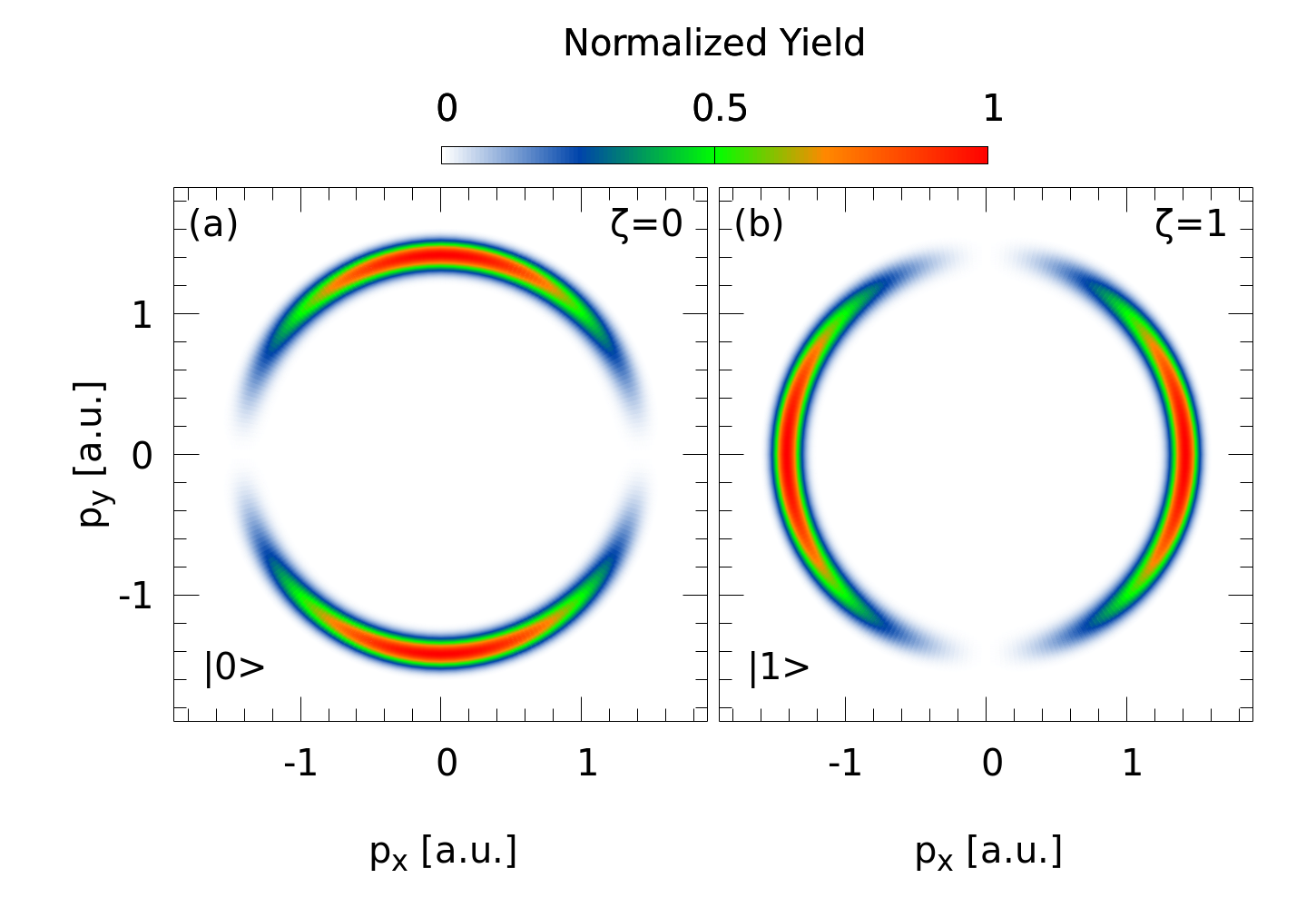}
	\caption{
PMD in the $p_x$–$p_y$ plane for a hydrogen atom driven by an ultrafast XUV pulse. The polarization mixing parameter $\zeta$ controls the redistribution of the field between orthogonal components. With $\zeta = 0$, emission is aligned along the $y$- axis, the associated continuum wavepacket is aliased as the state $\ket{0}$ (a); however, with $\zeta = 1$, emission is oriented along the $x$-axis, and the wavepacket is aliased as the state $\ket{1}$ (b).  
	}
	\label{fig:pmd}
\end{figure}
 
Two driving-field configurations are considered in this work, both exploiting polarization-mixing as the primary control mechanism. The first, hereafter referred to as the `polarization-mixing'  configuration, employs a monochromatic field. The second, hereafter referred to as the `bi-chromatic' configuration, introduces a second harmonic component alongside independent polarization control, providing access to additional ionization pathways and enabling coherent state engineering within an expanded four-dimensional photoelectron-state manifold. Having established the underlying interference and control mechanisms in the minimal two-state setting, the bichromatic configuration is then investigated as a natural route toward higher-dimensional continuum-state engineering. In both cases, the driving field $\mb{E}(t) = -\partial \mathbf{A}/\partial t$ is described in terms of the vector potential $\mb{A}(t) = A_x(t)\,\hat{x} + A_y(t)\,\hat{y}$.

The peak intensity \(I_0 = 1 \times 10^{14}~\mathrm{W/cm^2}\) primarily controls the contrast in the photoelectron momentum distributions; since normalized yields are plotted throughout, the observed angular structures are not sensitive to the precise value of the peak intensity for a given Keldysh parameter. For the polarization-mixing case, the carrier wavelength of \(30~\mathrm{nm}\) is considered. The orthogonal components are
\be
\left.
\begin{aligned}
A_x(t) &= A_0 f(t) \,\zeta \sin(\omega_0 t), \\
A_y(t) &= A_0 f(t) \,(1-\zeta)\cos(\omega_0 t + \varphi)
\end{aligned}
\right\}
\label{case1_eqns}
\ee
where \(A_0 = E_0/\omega_0\) is the peak vector-potential amplitude with \(\omega_0\) being the carrier frequency. The peak electric-field amplitude $E_0\ \text{[a.u.]} \simeq 5.342 \times 10^{-9} I_0$. The parameter \(\zeta\) (referred to hereafter as the polarization-mixing parameter) controls the relative strengths of the two components, and \(\varphi\) is the relative phase between the \(x\) and \(y\) components.

For the bichromatic case, \(\omega_0\) corresponds to a carrier wavelength of \(60~\mathrm{nm}\). The Cartesian components read
\be
\left.
\begin{aligned}
A_x(t) &= A_0 f(t) \Bigg[ \zeta_1 \sin(\omega_0 t) + \frac{\zeta_2}{2} \sin(2\omega_0 t) \Bigg], \\
A_y(t) &= A_0 f(t) \Bigg[ \zeta_3 \sin(\omega_0 t + \varphi_1) + \frac{\zeta_4}{2} \sin(2\omega_0 t + \varphi_2) \Bigg]
\end{aligned}
\right\}
\label{case2_eqns}
\ee
where the independent parameters \(\zeta_1, \zeta_2, \zeta_3,\) and \(\zeta_4\) control the relative strengths of the fundamental and second-harmonic components and thereby govern the population distribution among the dominant ionization pathways, while \(\varphi_1\) and \(\varphi_2\) provide independent control over the relative phases between the \(x\) and \(y\) components. The resulting photoelectron wave packets serve as the initial orthonormal basis vectors. The temporal envelope $f(t) = \sin^2(\pi t/\tau), \quad 0 \le t \le \tau$ is considered for both cases, with $\tau = 10$ optical cycles for respective wavelengths of 30 nm (polarization-mixing case) and 60 nm (bi-chromatic case). 

\section{Results and Discussion} \label{sec:results}

We now present the photoelectron momentum distributions (PMDs) obtained from the TDSE simulations for the two driving-field configurations introduced above. The first configuration explores the role of polarization mixing in shaping the angular emission patterns [Sec. \ref{sectionA}], while the second demonstrates coherent control of ionization pathways through a bichromatic field [Sec. \ref{sectionB}]. In both cases, emphasis is placed on the sensitivity of the normalized PMDs to the key control parameters, revealing clear signatures of interference and pathway manipulation in the strong-field ionization dynamics.


\subsection{Two-State Continuum Control}
\label{sectionA}

\begin{figure}[b]
	\includegraphics[width=\linewidth]{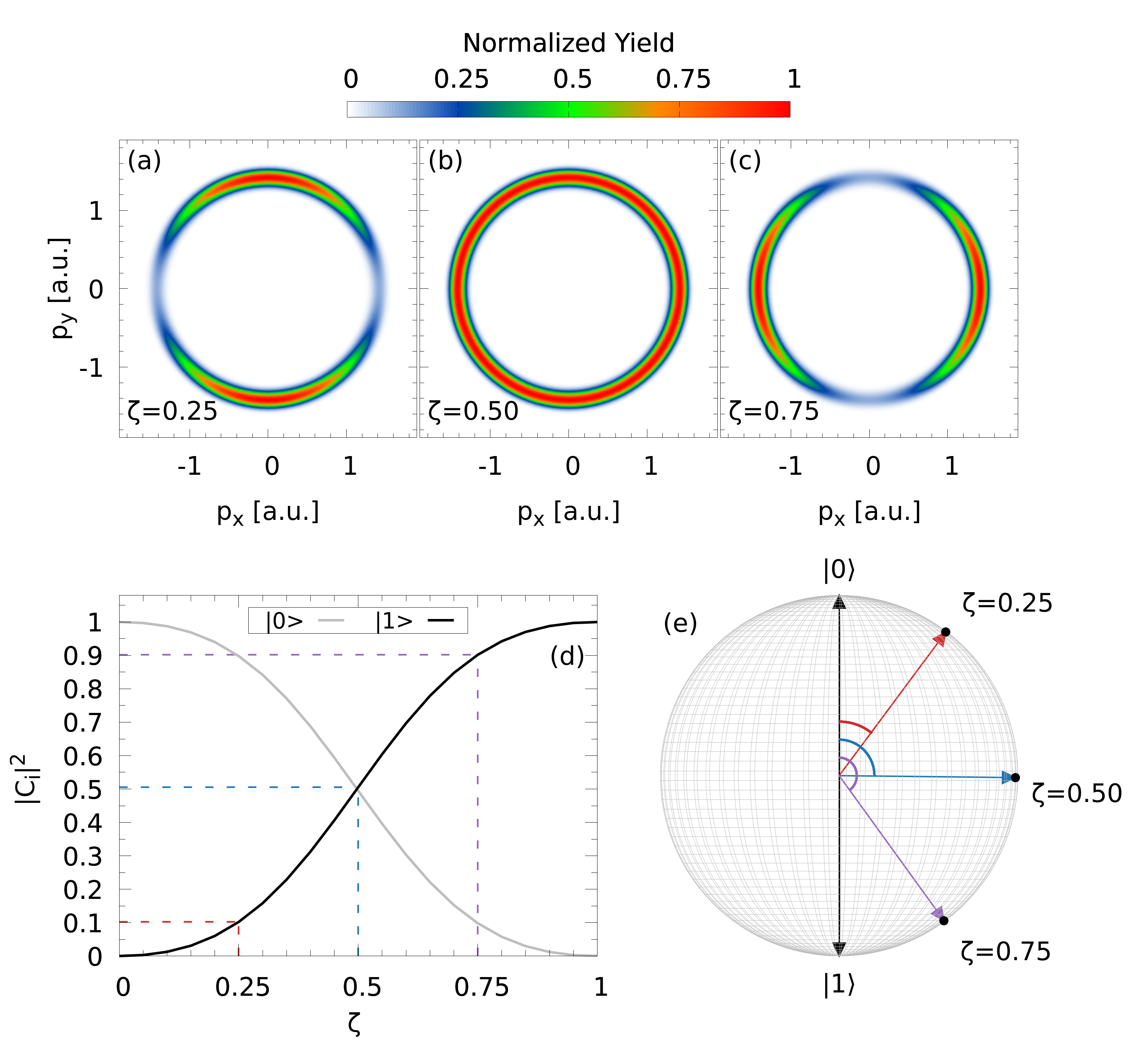}
	\caption{
	Panels (a)-(c) show the normalized PMD in the $p_x$-$p_y$ plane for $\zeta=0.25$, $0.50$, and $0.75$, respectively with $\varphi = 0$. 
		 The corresponding projection probabilities $|c_i|^2$ onto the effective basis states $\ket{0}$ and $\ket{1}$ as a function of $\zeta$ are shown in (d). The dashed vertical lines indicate the population values associated with the representative momentum distributions shown in panels (a)-(c). The corresponding mapping onto an effective Bloch sphere is presented in (e).
	}
	\label{fig2}
\end{figure}

Now, for the polarization-mixing case, the PMD shown in Fig.~\ref{fig:pmd} exhibits a clear, controllable restructuring of the photoelectron emission pattern, governed by the parameter $\zeta$. Within our vector-potential-based formulation, the driving field is decomposed into orthogonal components, with $\zeta$ continuously redistributing the field strength between the $x$- and $y$-directions [refer Eq. \eqref{case1_eqns}].
For $\zeta = 0$ [Fig.~\ref{fig:pmd}(a)], the field is fully polarized along the $y$-axis, resulting in a momentum distribution along $p_y$-direction, with two dominant lobes at positive and negative $p_y$. We aliased the associated continuum wavefunction in this configuration, i.e., $\zeta = 0$, as the state $\ket{0}$. In contrast, for $\zeta = 1$ [Fig.~\ref{fig:pmd}(b)], the field is fully polarized along the $x$-direction, leading to a rotation of the momentum distribution along the $p_x$-direction. The associated continuum wavefunction for this configuration is aliased as $\ket{1}$.
The states $\ket{0}$ and $\ket{1}$ thus form an \emph{effective two-state basis} in momentum space, defined by orthogonal emission directions. Intermediate values of $\zeta$ correspond to controlled mixtures of these basis states, providing a direct route toward preparing continuum electron wave packets in tunable superposition states and hence establishing $\zeta$ as a knob for steering and encoding information.

The orthogonality of the continuum states $\ket{0}$ and $\ket{1}$ is verified by explicitly evaluating their overlap in momentum space. The corresponding wave packets are first normalized according to $\int \psi(\mb{p})|^2\  d\mb{p} = 1$, and the overlap integral $\braket{0|1} \equiv \int \psi_0^\star(\mb{p}) \psi_1(\mb{p})\  d\mb{p}$ is computed on the same momentum grid using appropriate quadrature weights. The resulting projection matrix is found to be diagonal to numerical precision, with off-diagonal elements of the order of $10^{-8}$ and $|\braket{0|1}|^2 \sim 10^{-16}$. This near-vanishing overlap demonstrates that the two states are orthogonal within numerical accuracy. These results confirm that $\ket{0}$ and $\ket{1}$ constitute a well-defined orthonormal basis in the continuum, thereby validating their interpretation as an effective two-state system. Furthermore, the coherence properties of these two orthogonal states are discussed in Appendix-\ref{Appendix_A}.

Having established that $\ket{0}$ and $\ket{1}$ form an orthonormal continuum basis, the photoelectron wave packet generated for an arbitrary value of $\zeta$ can be expressed as,
\begin{equation}
	\ket{\psi}=c_0\ket{0}+c_1\ket{1},
\end{equation}
where the complex amplitudes $c_0 \equiv \braket{0|\psi}$ and $c_1\equiv \braket{1|\psi}$ are obtained by projection onto the basis states.
For a normalized wave packet, $|c_0|^2+|c_1|^2=1$.
To visualize the continuum state, we map it onto an effective Bloch sphere through the standard parameterization
\begin{equation}
	c_0=\cos\Big(\frac{\theta}{2}\Big) \text{e}^{i\phi_0},
	\qquad
	c_1=\sin\Big(\frac{\theta}{2}\Big) \text{e}^{i\phi_1},
\end{equation}
such that the physical state is completely specified by the populations
$|c_0|^2$ and $|c_1|^2$ together with the relative phase
\begin{equation}
	\Delta\phi=\phi_1-\phi_0=\arg(c_0^{\star}\ c_1).
	\label{relative_phase}
\end{equation}

In this representation, the mixing parameter $\zeta$ controls the population transfer between the basis states. It therefore determines the polar angle $\theta$, whereas the carrier-envelope phase controls the relative phase $\Delta\phi$ and hence the azimuthal angle on the Bloch sphere.
 
\begin{figure}[t]
	\includegraphics[width=\linewidth]{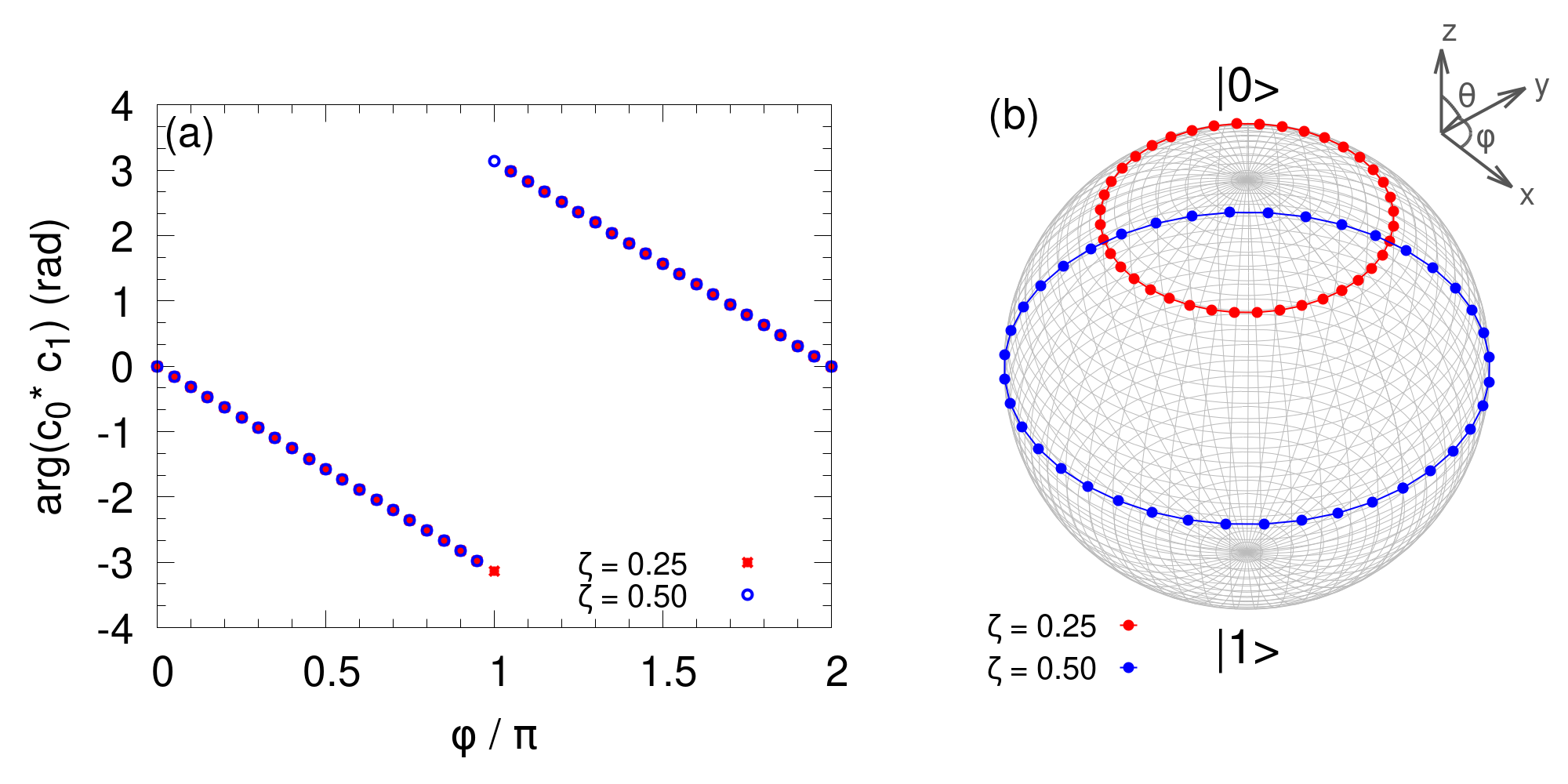}
	\caption{
		(a) Relative phase $\arg(c_0^{\star}c_1)$ as a function of the CEP $\varphi$ for $\zeta=0.25$ and $0.50$. The phase exhibits an approximately linear dependence on relative CEP, with the discontinuity near $\varphi=\pi$ arising from phase wrapping at $\pm\pi$. 
		(b) Corresponding trajectories on the effective Bloch sphere. The different values of $\zeta$ map onto distinct polar angles, while the relative CEP continuously controls the azimuthal evolution of the continuum quantum state along circles of constant latitude.
	}
	\label{fig3}
\end{figure}

 To further elucidate the role of control parameter $\zeta$, we examine the evolution of the  PMDs for intermediate values of $\zeta$, as shown in Fig.~\ref{fig2}(a-c) with $\varphi = 0$. While the limiting cases $\zeta = 0$ and $\zeta = 1$ define the two orthogonal basis states $\ket{0}$ and $\ket{1}$ (Fig.~\ref{fig:pmd}), intermediate $\zeta$ values produce distributions that cannot be described as purely aligned along either axis.
For $\zeta = 0.25$ [Fig.~\ref{fig2}(a)], the PMD remains predominantly oriented along the $p_y$ direction, retaining strong overlap with the $\ket{0}$ state, but already exhibits a noticeable redistribution toward the orthogonal $p_x$ channel. At $\zeta = 0.50$ [Fig.~\ref{fig2}(b)], the distribution becomes nearly isotropic along the emission ring, indicating an approximately equal contribution from both orthogonal emission directions. For $\zeta = 0.75$ [Fig.~\ref{fig2}(c)], the distribution is largely rotated toward the $p_x$ axis, approaching the $\ket{1}$ configuration.

To quantify this transition, the projection of the continuum states (resulting from different values of the $\zeta$) onto the two basis states $\ket{0}$ and $\ket{1}$ is carried out. The resulting populations are shown in Fig.~\ref{fig2}(d), which exhibit a clear, monotonic transfer of weight between the two states as $\zeta$ is varied. At $\zeta = 0$, the system resides entirely in $\ket{0}$, while at $\zeta = 1$ it is fully transferred to $\ket{1}$, consistent with the PMDs in Fig.~\ref{fig:pmd}. Importantly, at $\zeta = 0.5$, populations of the two states become equal, indicating a balanced contribution from the orthogonal emission channels. The nearly symmetric population exchange demonstrates that $\zeta$ provides direct control over the relative amplitudes of the effective continuum basis states through redistribution of the orthogonal field components.

The controlled projection on the basis states \(\ket{0}\) and \(\ket{1}\) can be visualized on an effective Bloch sphere, as shown in Fig.~\ref{fig2}(e). In this representation, the north and south poles correspond to \(\ket{0}\) and \(\ket{1}\), respectively. For a fixed relative phase \(\varphi\), intermediate values of the polarization-mixing parameter \(\zeta\) trace a continuous trajectory along a meridian (great circle passing through both poles). In particular, the case \(\varphi = 0\) corresponds to the prime meridian. The specific values \(\zeta = 0.25\), \(0.50\), and \(0.75\) mark distinct points along this path, illustrating a gradual rotation of the effective state vector. Notably, the point at \(\zeta = 0.50\) lies on the equatorial plane, consistent with the equal-weight superposition observed in Fig.~\ref{fig2}(d).

To complete the characterization of an effective two-state continuum system, we now examine the phase relation between the basis states  $\ket{0}$ and $\ket{1}$. While Fig.~\ref{fig2} establishes control over the relative populations, the relative phase is accessed by varying the relative CEP, $\varphi$, of the driving field (hereafter referred to as CEP only). Figure~\ref{fig3}(a) presents the evolution of the relative phase $\Delta \phi$ [Eq.~\eqref{relative_phase}] as a function of CEP for two representative values, $\zeta = 0.25$ and $\zeta = 0.50$. In both cases, the phase exhibits a clear linear dependence on CEP, spanning a full $2\pi$ range. This demonstrates that the CEP acts as a direct and robust control parameter for the phase difference between the two continuum states. Importantly, the slope of the phase evolution is identical for both $\zeta$ values, indicating that CEP-driven phase control is largely independent of the population distribution between the two basis states. This separation of roles, with $\zeta$ controlling amplitudes and CEP controlling phase, establishes independent tuning of the two key degrees of freedom that define the effective state.

The phase evolution is again visualized geometrically using the Bloch-sphere representation shown in Fig.~\ref{fig3}(b). For fixed $\zeta$, variation of CEP generates closed trajectories on the sphere corresponding to azimuthal rotations at a fixed polar angle. For $\zeta = 0.25$ (red), the trajectory forms a ring in the upper hemisphere, reflecting a state with a dominant $\ket{0}$ population and a CEP-dependent phase. For $\zeta = 0.50$ (blue), the trajectory lies on the equatorial plane 
(within the numerical accuracy), consistent with equal population of $\ket{0}$ and $\ket{1}$, while still exhibiting full azimuthal phase evolution. These circular trajectories demonstrate that CEP induces a controlled rotation of the effective state vector around the vertical axis of the Bloch sphere, while $\zeta$ determines its polar position. Together, they provide complete coverage of the accessible state space. The combined influence of the polarization-mixing parameter \(\zeta\) and the CEP \(\varphi\) is explored in Appendix~\ref{Appendix_B}, which shows that these two parameters can be tuned independently to realize continuum wave packets possessing any desired population ratio and relative phase.

\begin{figure}
	\includegraphics[width=\linewidth]{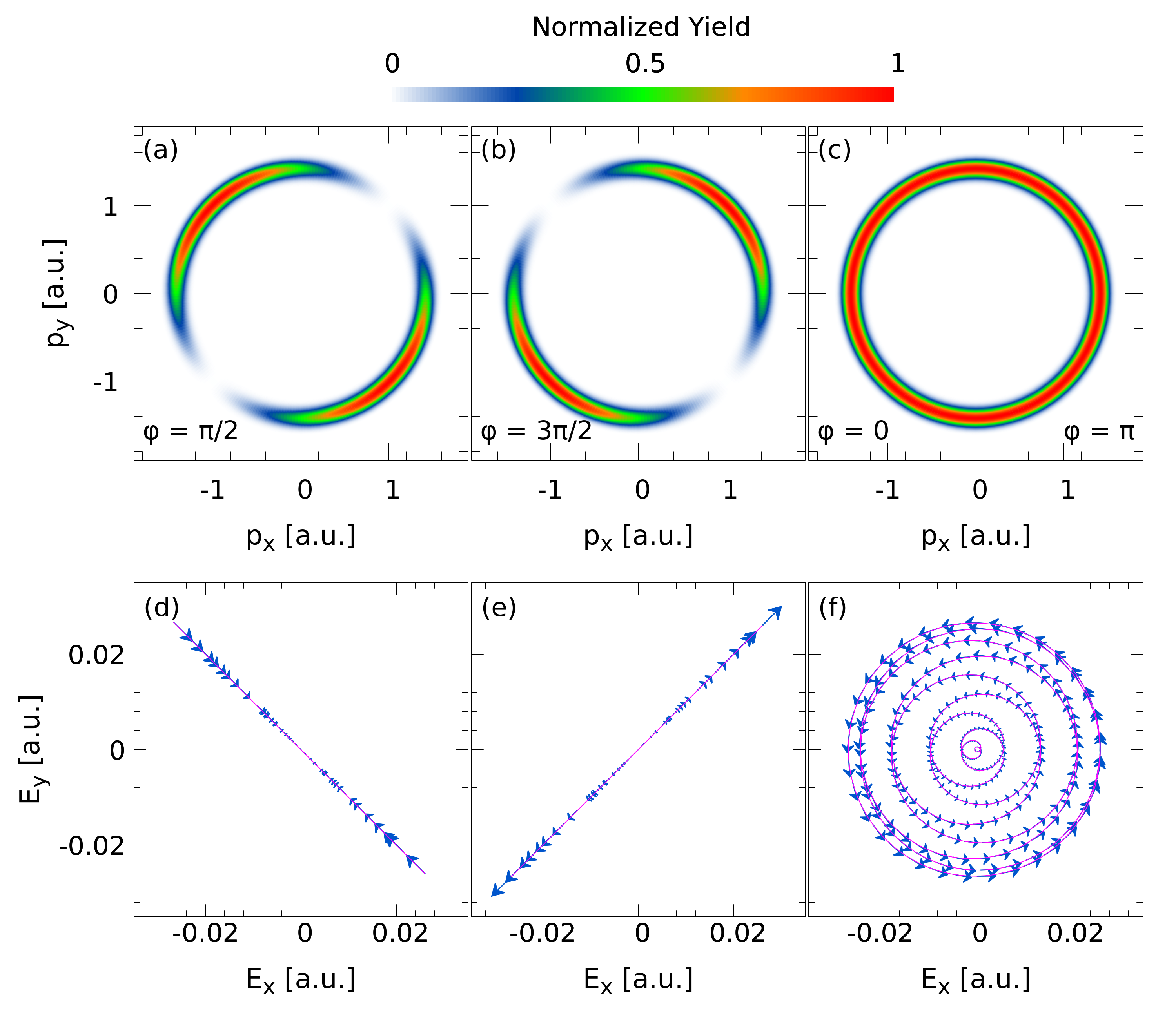}
	\caption{ Panels (a)-(c) show the normalized PMDs in the $p_x$-$p_y$ plane for $\varphi=\pi/2$, $3\pi/2$, and $\varphi=\pi$ (identical PMD for $\varphi=0$), respectively. However, panels (d)-(f) display the corresponding electric-field trajectories in the $E_x$-$E_y$ plane. The field trajectories for $\varphi=0$ and $\varphi=\pi$ are identical in shape; however, the direction of rotation is reversed for $\varphi=0$. Hence, the fields for only the $\varphi=\pi$ case are shown in panel (f). Here, the parameter $\zeta = 0.5$ is considered. 
	}	
	\label{fig4}
\end{figure}

To obtain microscopic insight into the CEP-dependent continuum dynamics, we present in Fig. \ref{fig4} the driving-field trajectory along with the resulting PMDs (for $\zeta = 0.5$). The upper panels [Fig.~\ref{fig4}(a-c)] show that the PMD undergoes a pronounced restructuring as the phase is varied from $\varphi=\pi/2$ to $\varphi=3\pi/2$ and $\varphi=\pi(\varphi=0)$. For $\varphi=0$ and $\varphi=\pi$, the distributions retain an almost rotationally symmetric ring-like structure, indicating nearly uniform angular emission into the continuum. In contrast, at $\varphi=\pi/2$ and $\varphi=3\pi/2$ the PMD becomes strongly localized into two crescent-shaped sectors, accompanied by pronounced suppression of emission over the remaining angular regions, revealing a substantial breaking of rotational symmetry in momentum space.

The origin of this behavior is evident from the corresponding electric-field trajectories shown in Fig.~\ref{fig4}(d-f). For $\varphi=0$ and $\varphi=\pi$, the field will be circularly-polarized, and thereby producing nearly isotropic PMDs. At $\varphi=\pi/2$ and $\varphi=3\pi/2$, however, the field trajectory becomes essentially linear, restricting the ionization dynamics predominantly to a narrower set of momentum directions. This transition from spiral to linear field geometry directly maps onto the observed angular localization of the continuum emission. The close correspondence between the field trajectory and the PMD structure indicates that the driving-field geometry governs the effective angular-gating mechanism for continuum electron emission.

\begin{figure}[b]
	\includegraphics[width=\linewidth]{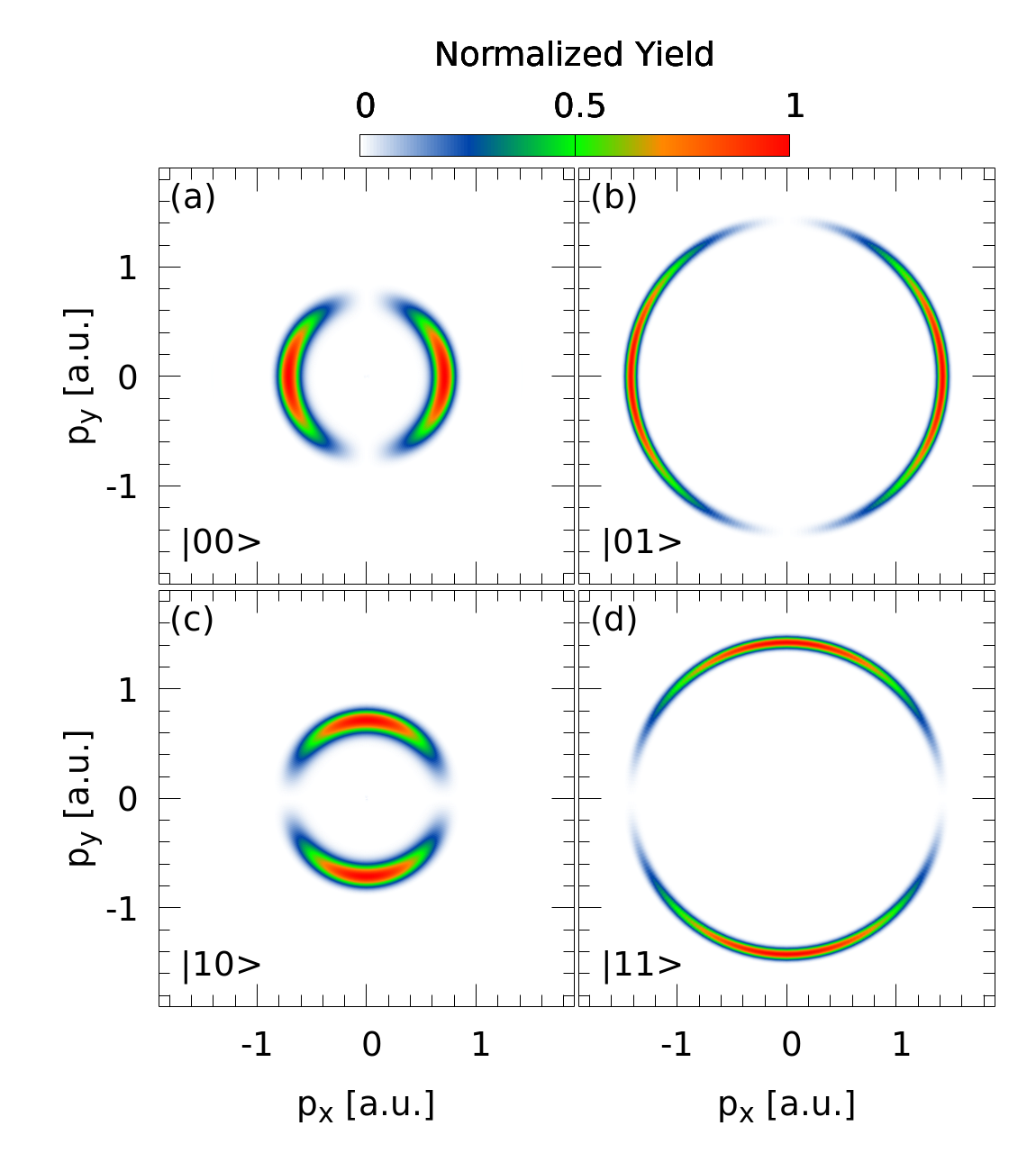}
	\caption{
PMDs in the $p_x$-$p_y$ plane defining the four effective continuum basis states. Panels (a)-(d) correspond to the states $\ket{00}$, $\ket{01}$, $\ket{10}$, and $\ket{11}$, respectively. 
	}
	\label{fig5}
\end{figure}

Taken together, Figs.~\ref{fig:pmd}-\ref{fig4} demonstrate that polarization-tailored XUV fields satisfy the essential requirements for coherent manipulation of an effective two-state continuum manifold. The basis states are orthogonal; their populations can be continuously tuned via the polarization-mixing parameter $(\zeta)$, and their relative phase can be independently controlled via the CEP. The resulting continuum dynamics remain directly observable through the corresponding PMDs, establishing a complete framework for state preparation and control within a momentum-space continuum basis.

\subsection{Multistate Continuum Control}
\label{sectionB}

Having established controllable preparation of an effective two-state continuum manifold, we now extend the scheme to a higher-dimensional continuum subspace composed of four orthogonal momentum-space basis states. This is done to demonstrate that the state-engineering protocol remains effective as the Hilbert-space dimension increases.  

The additional degree of freedom is introduced through more polarization-control parameters [Eq. \eqref{case2_eqns} with $\omega_0$ associated with the central wavelength of 60 nm], allowing independent steering of the angular emission structure in the continuum. This construction generates an effective four-dimensional continuum manifold that can be mapped onto the basis \(\{ \ket{00}, \ket{01}, \ket{10}, \ket{11} \}\). Explicitly, the manifold states corresponding to the parameter choices \(\zeta_i = \delta_{ij}\) (with \(i,j = 1,2,3,4\)) coincide with these basis states: 
setting \(\zeta_1 = 1\) and \(\zeta_{2,3,4} = 0\) yields \(\ket{00}\), 
\(\zeta_2 = 1\) and \(\zeta_{1,3,4} = 0\) yields \(\ket{01}\), and similarly for the remaining states.

\begin{figure}[t]
	\includegraphics[width=\linewidth]{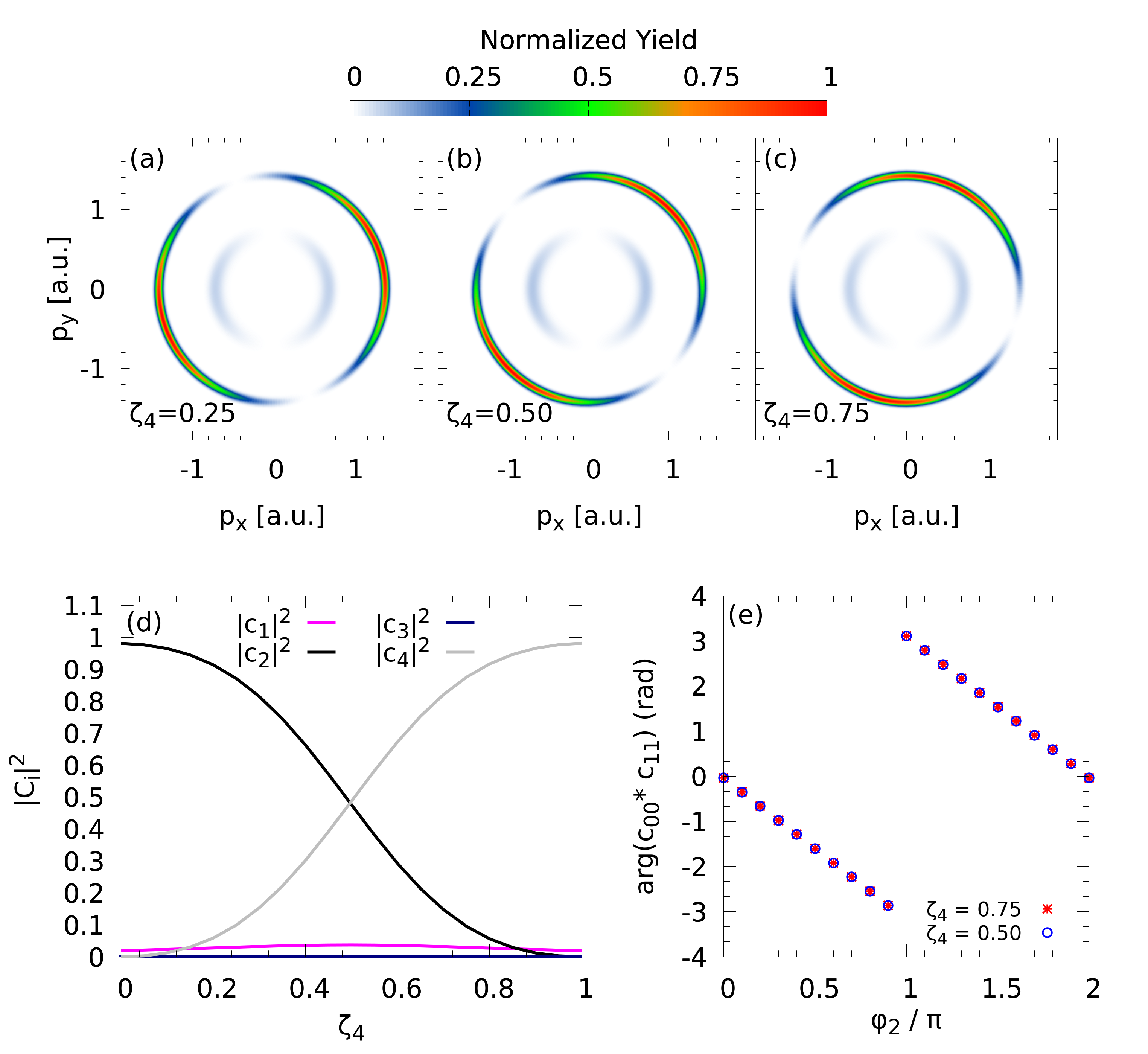}
	\caption{
	The normalized PMds for $\zeta_4=0.25$, $0.50$, and $0.75$, are presented in (a)-(c) respectively. However, the corresponding projection probabilities $|c_i|^2$ onto the four effective basis states as a function of $\zeta_4$ are shown in (d). The relative phase $\arg(c_{00}^{\star}c_{11})$ as a function of the CEP $\varphi_2$ with $\varphi_1 = 0$ for two representative values of $\zeta_4$ is presented in (e).
	}
	\label{fig6}
\end{figure}

The states \(\ket{00}\) (\(\zeta_1\)) and \(\ket{10}\) (\(\zeta_3\)) exhibit emission localized primarily along the \(p_x\) and \(p_y\) axes, respectively, at a radial momentum \(\approx \sqrt{2\omega_0 - 2 I_p}\). 
Conversely, the states \(\ket{01}\) (\(\zeta_2\)) and \(\ket{11}\) (\(\zeta_4\)) appear as momentum rings of radius \(\approx \sqrt{4\omega_0 - 2 I_p}\) with similar angular distributions. 
These two distinct radial manifolds therefore provide an additional degree of freedom beyond the angular emission channel, enabling the construction of a complete four-state continuum basis. The PMDs associated with these four-state continuum bases are presented in Fig. \ref{fig5}(a-d).

The corresponding PMDs exhibit negligible spatial overlap, ensuring clear distinguishability among the four configurations in momentum space [Fig.~\ref{fig5}(a-d)]. Similar to the effective two-state continuum case, the orthogonality stems from the combined angular and radial separation of the continuum wave packets. 
We quantified this by evaluating the overlap matrix \(S_{ij,kl}=\langle ij|kl\rangle\) via the momentum-space inner product. 
The matrix is diagonal, with off-diagonal elements satisfying \(|\langle ij|kl\rangle|^2\lesssim 10^{-16}\) for distinct states \((ij)\neq(kl)\). This demonstrates that the four states \(\{\ket{00},\ket{01},\ket{10},\ket{11}\}\) form a numerically orthogonal basis of excellent accuracy, constituting an effective four-dimensional Hilbert space encoded fully within the continuum electron momentum distribution.

Having established the orthonormal basis ${\ket{00},\ket{01},\ket{10},\ket{11}}$, an arbitrary continuum wave packet within this effective four-state manifold can be expressed as
\be \ket{\psi} = c_{00}\ket{00} + c_{01}\ket{01} + c_{10}\ket{10} + c_{11}\ket{11}, \ee
where the complex amplitudes $c_{ij}=\braket{ij|\psi}$ are obtained by projection onto the corresponding basis states and satisfy $\sum_{ij}|c_{ij}|^2=1$. The populations $|c_{ij}|^2$ characterize the occupation of the four continuum basis states. In contrast, coherence between different basis components is quantified by the relative phases, as discussed in Appendix-\ref{Appendix_A}. In the following, we use these populations and relative phases to characterize coherent control within the effective four-state continuum manifold.

To investigate coherent control within the effective four-state continuum, we fix \(\zeta_1 = 0.05\) and \(\zeta_3 = 0\), while treating \(\zeta_4\) as the primary control parameter and setting \(\zeta_2 = 1 - \zeta_4\). 
Any other combination of these parameters yields qualitatively similar behavior. The corresponding PMDs for these parameter values are presented in Fig.~\ref{fig6}(a-c). 

\begin{figure}[b]
	\includegraphics[width=\linewidth]{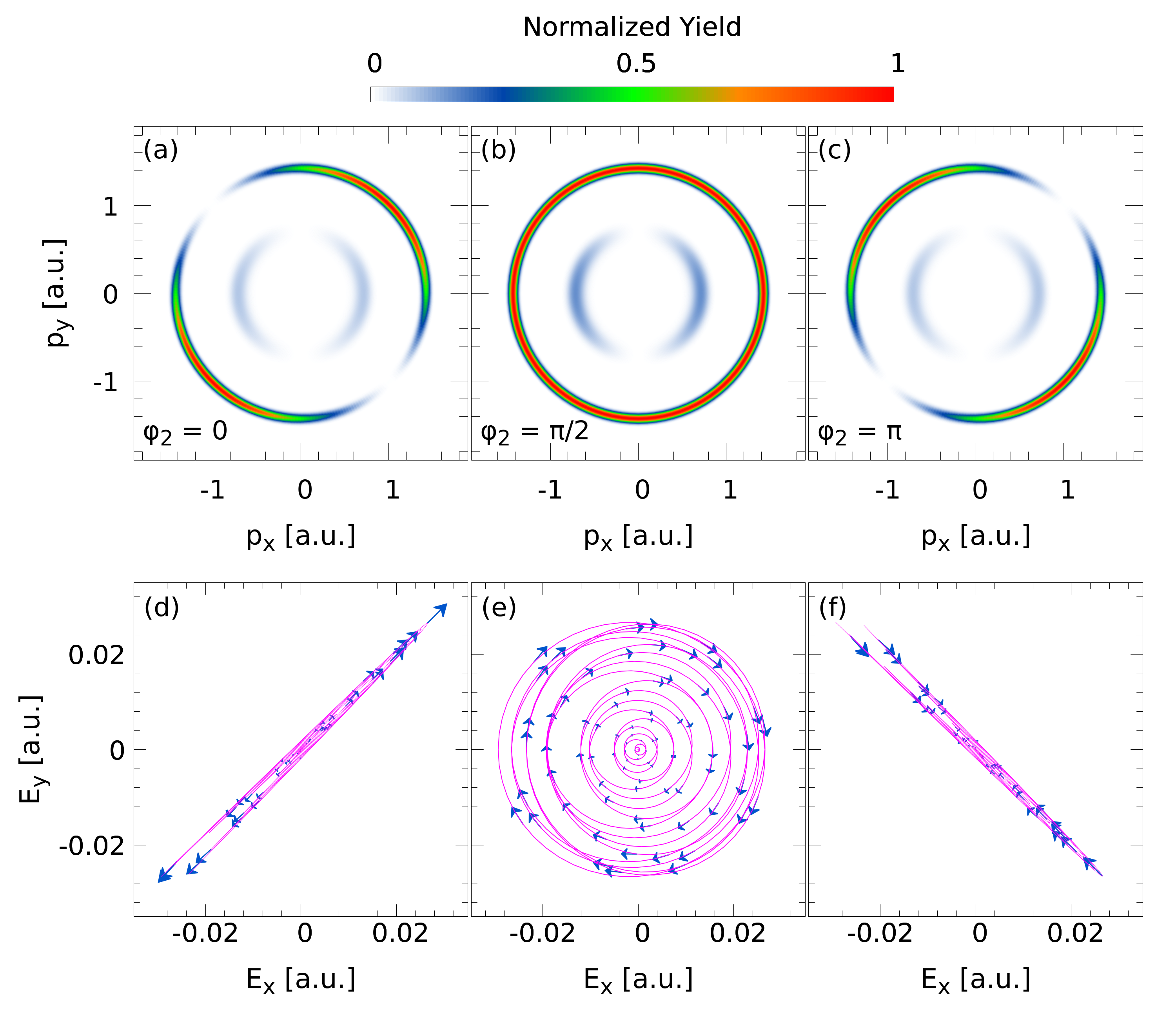}
	\caption{
		 The normalized PMDs for $\varphi_2=0$, $\pi/2$, and $\pi$, are presented respectively in (a)-(c). However, panels (d)-(f) display the corresponding electric-field trajectories in the $E_x$-$E_y$ plane. For all the cases, $\varphi_1 = 0$ is considered. Here, we have fixed  $\zeta_1 = 0.05, \zeta_2 = 0.5, \zeta_3 = 0$, and  $\zeta_4 = 0.5$.
	}
	\label{fig7}
\end{figure}

In contrast to the two-state case, the momentum distributions now contain contributions from both radial manifolds, reflecting the higher-dimensional structure of the continuum basis. As \(\zeta_4\) is varied from \(0.25\) to \(0.75\), the dominant angular emission sectors rotate continuously around the momentum ring, while the inner and outer radial components remain simultaneously populated. This evolution demonstrates that \(\zeta_4\) directly controls the composition of the continuum state.

To quantify the dynamics, the PMDs are projected onto the orthonormal basis \(\{ \ket{00}, \ket{01}, \ket{10}, \ket{11} \}\), with the resulting populations shown in Fig.~\ref{fig6}(d). For the field parameters considered here, the dynamics are dominated by population transfer between \(\ket{01}\) and \(\ket{11}\), while the remaining basis states retain only weak occupation. As \(\zeta_4\) increases from \(0\) to \(1\), population is transferred smoothly from \(\ket{01}\) to \(\ket{11}\), reaching approximately equal contributions near \(\zeta_4 = 0.5\). Note, however, that unlike the two-state scenario, the individual populations never reach unity, since other states remain weakly occupied throughout evolution due to $\zeta_1 = 0.05$. This behavior reflects the multi-dimensional coupling inherent to the four-state manifold. The nearly symmetric exchange nonetheless demonstrates coherent tunability of the continuum-state amplitudes through polarization engineering.

The corresponding phase dynamics are shown in Fig.~\ref{fig6}(e), where the relative phase $\phi_{11}=\arg(c_{00}^{\star}c_{11})$ is plotted as a function of the CEP $\varphi_2$ with $\varphi_1 = 0$. The remaining relative phases exhibit qualitatively similar behavior. For both representative values of $\zeta_4$, the phase depends approximately linearly on CEP over a full $2\pi$ interval, with the discontinuity near $\varphi_2=\pi$ arising from phase wrapping at $\pm\pi$. The nearly identical phase evolution for different values of $\zeta_4$ indicates that CEP-driven phase control remains largely independent of the population-transfer dynamics. Consequently, parameters $\zeta_i$ primarily control the continuum-state amplitudes, whereas $\varphi_i$ governs their relative phases. Together, these results demonstrate independent control of amplitude and phase within the effective four-state continuum manifold.

With \(\zeta_1 = 0.05\), \(\zeta_2 = 0.5\), \(\zeta_3 = 0\), and \(\zeta_4 = 0.5\) fixed, we now vary the carrier-envelope phase \(\varphi_2\) while keeping \(\varphi_1 = 0\). 
The Fig.~\ref{fig7} provides a microscopic interpretation of the CEP-induced control within the four-state continuum manifold by directly correlating the driving-field geometry with the resulting PMDs. The upper panels [Fig.~\ref{fig7}(a-c)] demonstrate that changing \(\varphi_2\) primarily modulates the angular distribution of the photoelectron emission while leaving the radial structure essentially unchanged. For \(\varphi_2 = 0\) and \(\varphi_2 = \pi\), the PMDs display localized crescent-shaped emission sectors concentrated along specific angular directions. In contrast, at \(\varphi_2 = \pi/2\) the emission becomes nearly rotationally symmetric, yielding an almost uniform distribution around the momentum ring.

The origin of this behavior is revealed by the corresponding electric-field trajectories shown in Fig.~\ref{fig7}(d-f). For $\varphi_2=0$ and $\varphi_2=\pi$, the field is predominantly linear. It should be noted that these are bi-chromatic fields and, hence, are not exactly linear, as seen in Fig. \ref{fig4}(d) and (e). The polarization of the driving fields favors ionization into a restricted set of momentum-space channels, thereby generating localized angular emission patterns. At $\varphi_2=\pi/2$, the field acquires a spiral-like structure that samples a broader range of emission directions, leading to a nearly isotropic continuum distribution. The close correspondence between the field geometry and the PMD structure demonstrates that CEP control operates by directly modifying the accessible continuum emission channels. These results identify polarization-tailored XUV fields as an effective tool for steering continuum electron dynamics within higher-dimensional momentum-space manifolds.

Taken together, Figs.~\ref{fig5}-\ref{fig7} demonstrate coherent control within an effective four-dimensional continuum manifold. The basis states are orthogonal and distinguishable by their angular- and radial-momentum-space structures. At the same time, the polarization-mixing parameter $\zeta_4$ and the CEP $\varphi_2$ provide independent control of the continuum-state amplitudes and relative phases. The resulting continuum dynamics remain directly observable through the corresponding PMDs, thereby establishing a scalable framework for coherent-state manipulation in higher-dimensional momentum-space manifolds.

\section{Conclusion} \label{sec:conclusion}

We have demonstrated coherent control of continuum electron states generated by polarization-tailored XUV fields through direct manipulation of photoelectron momentum distributions. By combining polarization mixing with CEP control, we constructed orthogonal momentum-space basis states and achieved independent control of their amplitudes and relative phases. Within an effective two-state continuum manifold, the polarization-mixing parameter governs population transfer between orthogonal basis states, while the CEP provides independent control of their relative phase, enabling coherent preparation and steering of continuum superposition states.

We further extended this framework to a four-dimensional continuum manifold spanned by orthogonal momentum-space states distinguished by both their angular emission patterns and radial momentum distributions. The resulting multistate continuum dynamics exhibit independent tunability of amplitudes and phases through experimentally accessible field parameters. Analysis of the corresponding electric-field trajectories and PMDs reveals that control arises from direct manipulation of the accessible continuum emission channels through the driving-field geometry, establishing polarization-tailored XUV fields as a versatile platform for engineering low-dimensional coherent state spaces within the photoelectron continuum.

We note that the present four-state basis does not exhibit entanglement and therefore does not constitute a genuine two-qubit system; the demonstrated results concern coherent control, population-phase tunability, and Bloch-sphere dynamics within a single-particle continuum manifold. Nevertheless, the PMD framework is inherently scalable: multicolor XUV fields can generate additional concentric momentum rings, each corresponding to a distinct energy manifold, providing a natural route toward higher-dimensional state spaces. Engineering multicolor driving fields whose angular-momentum subspaces support well-defined entangled qubits or qudits represents a promising direction for future work. Realizing this potential experimentally will require careful attention to practical decoherence sources, including CEP jitter, focal-volume averaging over intensity distributions, and finite detector momentum resolution, all of which affect the visibility of the coherent interference demonstrated here.

\begin{figure}[t]
\centering
\includegraphics[width=1.0\columnwidth]{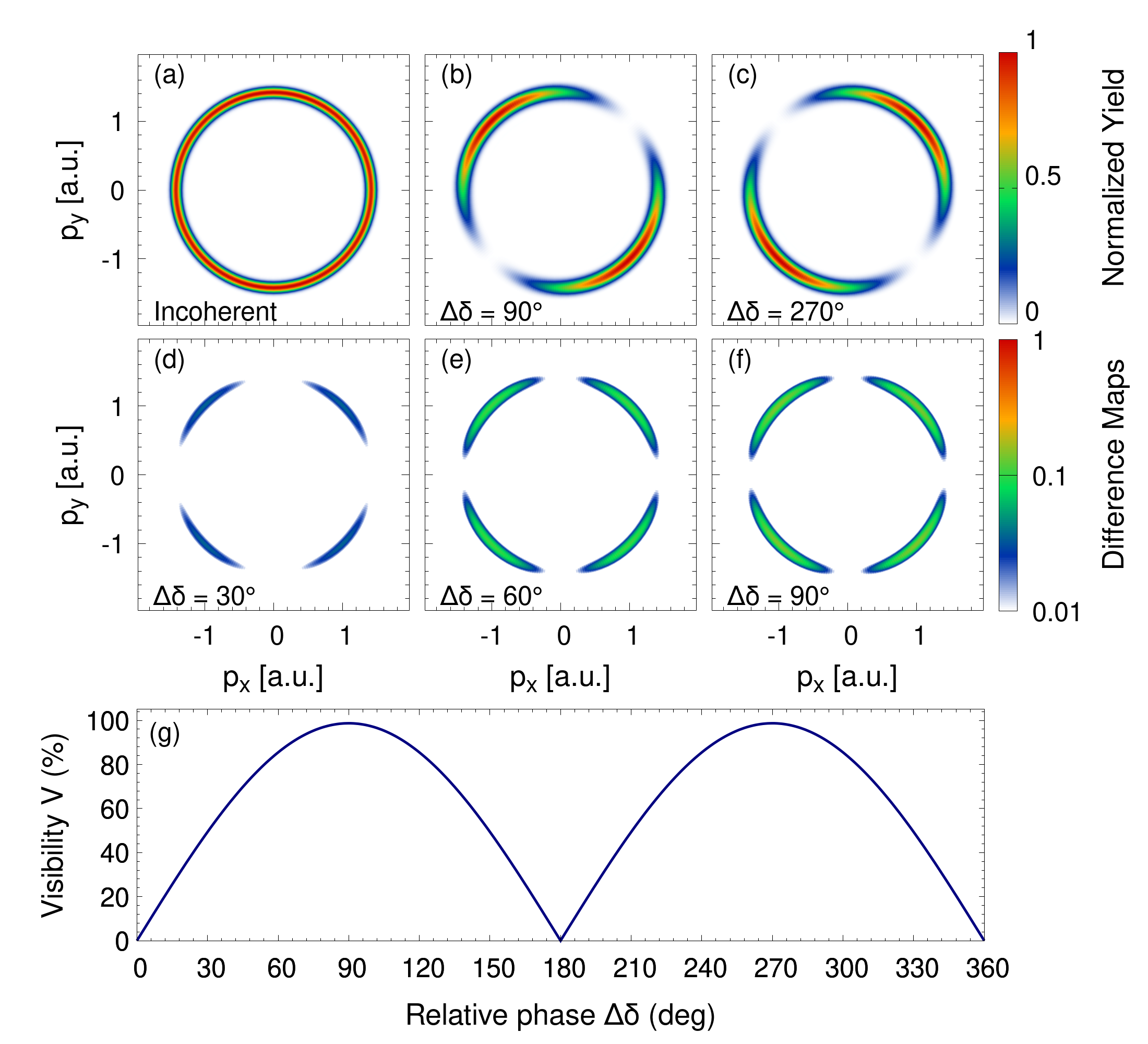}
\caption{Quantitative demonstration of quantum coherence at $\zeta=0.5$. 
 The PMDs for (a) Incoherent mixture, (b) coherent superposition with $\Delta\delta=90^\circ$, (c) coherent superposition with $\Delta\delta=270^\circ$. However, the difference maps [Eq. \ref{diff_maps}] for (d) $\Delta\delta=30^\circ$, (e) $\Delta\delta=60^\circ$, and (f) $\Delta\delta=90^\circ$ are shown.  Maximum interference visibility $V(\Delta\delta)$ as a function of relative phase is presented in (g). }
\label{fig:coherence}
\end{figure}

\section*{Acknowledgments} 
The authors acknowledge the Department of Science and Technology (DST) for providing computational resources through the FIST program (Project No. SR/FST/PS-1/2017/30). Also, the authors acknowledge BITS - Pilani, Pilani Campus, for providing the `Jayant' HPC facility. 

\section*{Data Availability}

The data that support the findings of this article are not publicly available. The data are available from the authors upon reasonable request.

\appendix

\section{Demonstration of Quantum Coherence}
\label{Appendix_A}
 To rigorously establish the presence of genuine quantum coherence between the orthogonal continuum states $|0\rangle$ and $|1\rangle$, we explicitly construct coherent superpositions and compare them with the corresponding incoherent statistical mixture at equal population ($\zeta=0.5$).

The ``coherent superposition'' with a controllable relative phase $\Delta\delta$ is given by
\be
|\psi_{\rm coh}(\Delta\delta)\rangle = \frac{1}{\sqrt{2}} \left( |0\rangle + \text{e}^{i \Delta\delta} |1\rangle \right).
\label{coh_mix}
\ee
The associated PMD is calculated as:
\be
{\rm PMD}_{\rm coh}(\mb{p};\Delta\delta) = \left| \frac{\psi_0(\mb{p}) + \text{e}^{i \Delta\delta} \psi_1(\mb{p})}{\sqrt{2}} \right|^2.
\label{coh_mix_pmd}
\ee

\begin{figure}[b]
	\centering
	\includegraphics[width=\linewidth]{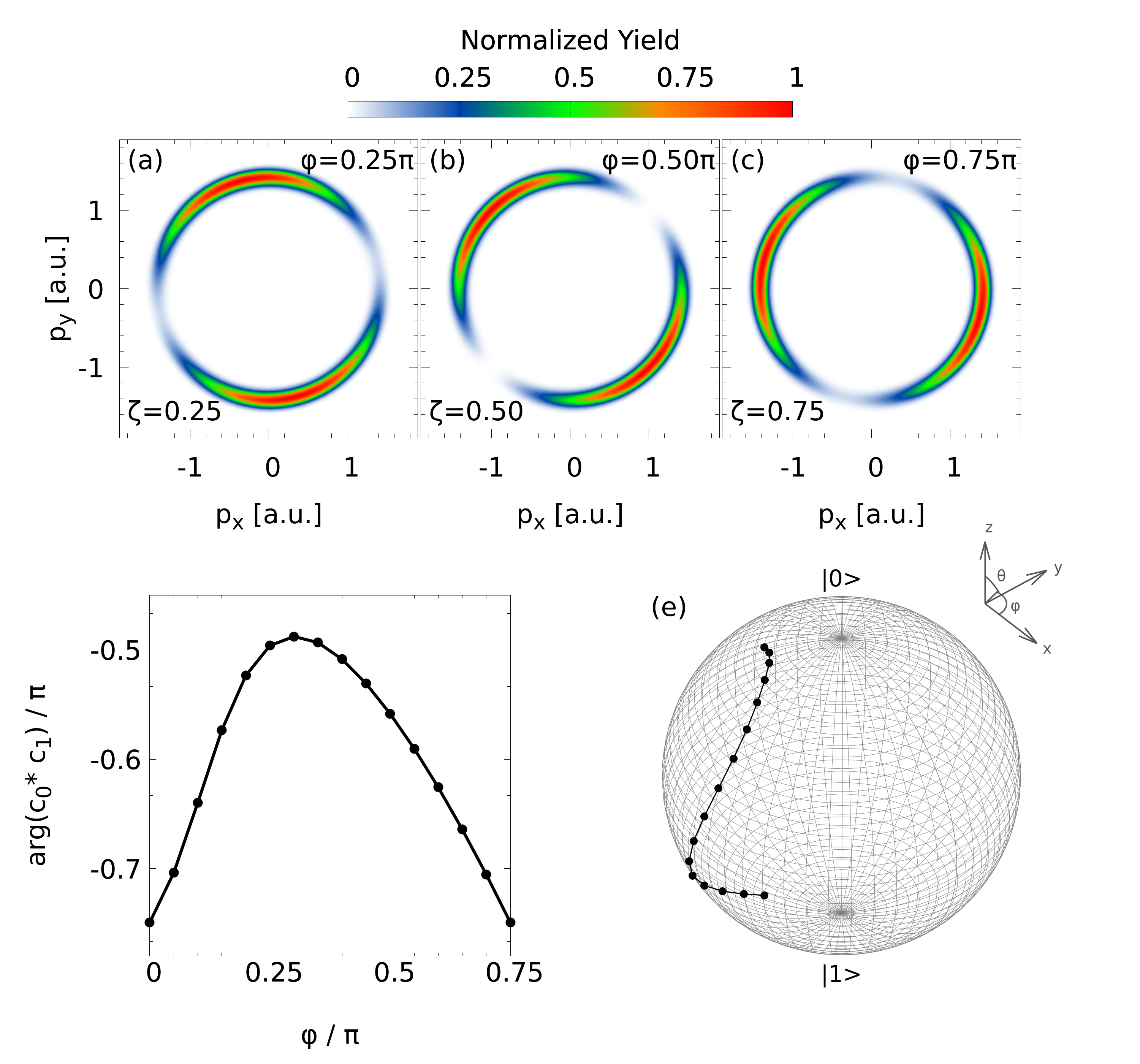}
	\caption{
		Panels (a)-(c) show normalized PMDs for $\zeta=0.25$, $0.50$, and $0.75$, with corresponding CEP values $\varphi=0.25\pi$, $0.50\pi$, and $0.75\pi$, respectively. (d) Relative phase $\arg(c_0^{\star}c_1)/\pi$ as a function of $\varphi$ for $\zeta\in[0.25,0.75]$. (e) Bloch-sphere representation of the resulting continuum states, where the trajectory is generated by simultaneously varying the polarization control parameter and CEP according to $\zeta=g$ and $\varphi = g\ \pi;\ g \in [0.25, 0.75]$ .
			}
	\label{fig8}
\end{figure}

In contrast, the ``incoherent statistical mixture'' (a classical mixture with no definite relative phase) corresponds to the density operator
\be
\hat{\rho}_{\rm incoh} = \frac{1}{2} |0\rangle\langle0| + \frac{1}{2} |1\rangle\langle1|,
\label{incoh_mix}
\ee
which yields the PMD given as:
\be
{\rm PMD}_{\rm incoh}(\mb{p}) = \frac{1}{2} \left( |\psi_0(\mb{p})|^2 + |\psi_1(\mb{p})|^2 \right).
\label{incoh_mix_pmd}
\ee

The difference between the coherent and incoherent PMDs is solely due to the ``interference cross term''
\be
{\rm PMD}_{\rm coh}(\mb{p};\Delta\delta) - {\rm PMD}_{\rm incoh}(\mb{p}) = {\rm Re}\left[\text{e}^{i \Delta\delta} \psi_0^\star(\mb{p}) \psi_1(\mb{p}) \right].
\label{diff_maps}
\ee
Although the two basis states are globally orthogonal ($\langle0|1\rangle \approx 0$), their local overlap $\psi_0^\star(\mb{p})\psi_1(\mb{p})$ is finite in specific angular regions of momentum space. This local overlap enables observable quantum interference when a well-defined relative phase is present.

We define the ``maximum interference visibility'' as:
\be
V(\Delta\delta) = \frac{\max_{\mb{p}} \left| {\rm PMD}_{\rm coh}(\mb{p};\Delta\delta) - {\rm PMD}_{\rm incoh}(\mb{p}) \right|}{\max_{\mb{p}} [{\rm PMD}_{\rm incoh}(\mb{p})]} \times 100\%.
\label{visibility}
\ee
This quantity measures the relative strength of the interference contribution with respect to the incoherent background.

The representative PMDs: the incoherent mixture [Eq. \eqref{incoh_mix_pmd}, nearly isotropic ring] and coherent superpositions [Eq. \eqref{coh_mix_pmd}] at $\Delta\delta=90^\circ$ and $270^\circ$ are presented in Fig. \ref{fig:coherence}(a-c). However, the difference maps [Eq. \ref{diff_maps}] at $\Delta\delta=30^\circ$, $60^\circ$, and $90^\circ$ are shown in Fig. \ref{fig:coherence}(d-f) respectively. These difference plots directly isolate the phase-dependent interference pattern, whose angular position and amplitude vary systematically with $\Delta\delta$. Moreover, Fig. \ref{fig:coherence}(g) shows the modulation of the ``maximum interference visibility'' given by Eq. \eqref{visibility} with the relative phase difference $\Delta\delta$. The visibility reaches a maximum of nearly 99\% near $\Delta\delta = 90^\circ$ and exhibits a clear sinusoidal dependence on the relative phase.

It is important to emphasize that the phase $\Delta\delta$ in the above analysis was imposed manually after extracting the basis states. An equivalent control over the relative phase is achieved `physically' by varying the carrier-envelope phase (CEP) of the driving XUV field, as shown in Fig. \ref{fig3} and Fig. \ref{fig4}. The excellent consistency between the manual phase scan and the CEP-driven results confirms that the CEP provides direct and robust experimental access to the relative phase between the orthogonal continuum states, as can also be corroborated by comparing Fig. \ref{fig4}(a)-(b) with Fig. \ref{fig:coherence}(b)-(c). 

Taken together, the high visibility, clear interference patterns in the difference maps, and the direct CEP control demonstrate that $|0\rangle$ and $|1\rangle$ form a genuine coherent two-level system embedded in the photoelectron continuum.

\section{Joint variation of $\zeta$ and $\varphi$}
\label{Appendix_B}

Figure~\ref{fig8} illustrates the evolution of the two-state continuum manifold when the polarization control parameter $\zeta$ and the CEP $\varphi$ are varied simultaneously. Unlike the single-parameter cases discussed in the main text, where either the state populations or the relative phase can be controlled in a comparatively straightforward manner, the joint variation of $\zeta$ and $\varphi$ produces a highly coupled evolution of the continuum state. As evident from the PMDs in Figs.~\ref{fig8}(a)-(c), changes in the CEP not only rotate and redistribute the momentum distributions but also modify the relative contributions of the two continuum basis states through their dependence on $\zeta$. Consequently, both the population amplitudes and the phase coherence evolve simultaneously.
 
The relative phase $\arg(c_0^\star c_1)$ shown in Fig.~\ref{fig8} exhibits a non-monotonic dependence on $\phi$, indicating that the phase evolution cannot be inferred independently of the accompanying population dynamics. This coupling is further visualized on the Bloch sphere in Fig.~\ref{fig8}(e), where the trajectory generated by simultaneously varying $\zeta = g$ and $\varphi = g\ \pi;\ g \in [0.25, 0.75]$ follows a curved path rather than a simple meridional or azimuthal rotation. This behavior arises because the two control parameters, although independent in principle, act on different degrees of freedom. The parameter $\zeta$ primarily modifies the population distribution between the basis states. At the same time, $\varphi$ influences their relative phase. When both parameters are varied together, the combined effect produces a trajectory on the Bloch sphere that cannot be decomposed into independent contributions from each parameter alone. The figure highlights the complexity of multi-dimensional coherent control in continuum-state manifolds. It demonstrates that clean, independent tuning of population and relative phase is generally not possible with these field parameters, even though the controls themselves can be adjusted independently.

\bibliographystyle{apsrev4-2}

%

\end{document}